\documentclass[aps,showpacs,nofootinbib,preprintnumbers]{revtex4}

\usepackage{graphicx}
\usepackage{graphics}
\usepackage{amssymb}
\usepackage{amsmath}

\newcommand{\sig}{\sigma}
\newcommand{\eps}{\epsilon}

\newcommand{\be}{\begin{equation}}
\newcommand{\ee}{\end{equation}}
\newcommand{\bea}{\begin{eqnarray}}
\newcommand{\eea}{\end{eqnarray}}

\newcommand{\nslash}{\kern 0.2 em n\kern -0.50em /}
\newcommand{\kslash}{\kern 0.2 em k\kern -0.45em /}
\newcommand{\pslash}{\kern 0.2 em p\kern -0.50em /}
\newcommand{\Sslash}{\kern 0.2 em S\kern -0.50em /}
\newcommand{\Pslash}{\kern 0.2 em P\kern -0.50em /}
\newcommand{\Rslash}{\kern 0.2 em R\kern -0.50em /}
\newcommand{\open}{{<\kern -0.3 em{\scriptscriptstyle )}}}

\newcommand{\sT}{T}

\begin{document}
\title{Evolution equations for extended dihadron fragmentation functions}

\author{Federico A. Ceccopieri}
\email{ceccopieri@fis.unipr.it}
\affiliation{Dipartimento di Fisica, Universit\`a di Parma,\\
Viale delle Scienze, Campus Sud, 43100 Parma, Italy}

\author{Marco Radici}
\email{marco.radici@pv.infn.it}
\affiliation{Dipartimento di Fisica Nucleare e Teorica, Universit\`{a} di Pavia, 
and\\
Istituto Nazionale di Fisica Nucleare, Sezione di Pavia, I-27100 Pavia, Italy}

\author{Alessandro Bacchetta}
\email{alessandro.bacchetta@desy.de}
\affiliation{Theory Group, Deutsches Elektronen-Synchroton DESY, \\
D-22603 Hamburg, Germany}

\begin{abstract}
We consider dihadron fragmentation functions, describing the fragmentation 
of a parton in two unpolarized hadrons, and in particular
extended dihadron fragmentation functions, explicitly dependent on the
invariant mass,  $M_h$, of the hadron pair. We first rederive the known results
on $M_h$-integrated functions using Jet Calculus
techniques, and then we present the evolution equations for extended dihadron
fragmentation functions. Our results are relevant for the analysis of experimental
measurements of two-particle-inclusive processes at different energies.
\end{abstract}


\preprint{DESY 07-034}

\maketitle

\section{Introduction}
\label{sec:intro}

The fragmentation of partons into hadrons
has been studied in detail in semi-inclusive
processes with one hadron detected in the final state, such as 
$e^+ e^-$ annihilation, Semi-Inclusive Deep-Inelastic Scattering (SIDIS) or
hadron-hadron collisions. Factorization 
theorems (see, e.g., \cite{Ellis:1978ty,Amati:1978wx,Collins:1988gx}) allow to 
separate perturbatively calculable short-range coefficient 
functions from long-distance nonperturbative fragmentation functions 
$D_1^{i\to h}(z)$, describing the ``decay" of the hard parton $i$ into an observed 
hadron $h$ with fractional energy $z$, provided a hard scale is available. 
This is the case of the $e^+ e^- \to h X$ 
reaction, where the coefficient function is known at least to 
$O(\alpha_s)$~\cite{Altarelli:1979kv,Furmanski:1981cw}, with $\alpha_s$ the
running strong coupling constant.
The same fragmentation functions, $D_1^{i\to h}$, occur 
in the factorized formula for SIDIS at $O(\alpha_s)$~\cite{Altarelli:1979kv,Furmanski:1981cw},
combined with the specific process-dependent coefficient functions at the same
accuracy. For hadronic collisions, factorization is usually assumed, but has not been
proven yet. 

When considering semi-inclusive processes with two detected hadrons in the
final state, e.g., $e^+ e^- \to h_1 h_2 X$,
a new class of fragmentation functions, the so-called Dihadron Fragmentation
Functions (DiFF), needs to be introduced to guarantee factorization of all collinear
singularities~\cite{Konishi:1978yx}. From this perspective, DiFF are analogous
to fracture functions in the space-like 
regime~\cite{Trentadue:1993ka,Graudenz:1994dq}.
The DiFF evolution equations have been recently reanalyzed in Ref.
~\cite{Majumder:2004wh} and 
the cross section for $e^+ e^- \to h_1 h_2 X$ has been calculated to $O(\alpha_s)$ 
in Ref.~\cite{deFlorian:2003cg}.
At $O(\alpha_s^0)$, the production  of two hadrons $h_1$, $h_2$ 
with fractional energies $z_1$, $z_2$ and belonging to the same jet, is described by
a DiFF, $D_1^{i\to h_1 h_2}(z_1,z_2)$, i.e., the fragmentation of a
single parton $i$ into the two hadrons.  At $O(\alpha_s)$,  
hadrons produced in the same jet could either come from the fragmentation of a
single parton into two hadrons or by the fragmentation of two collinear
partons, $i$ and  $j$, into single hadrons.
This implies that evolution equations for DiFF contain an inhomogeneous term of
the form $D_1^{i\to h_1} \otimes D_1^{j\to h_2}$~\cite{deFlorian:2003cg}.

All these studies focused on DiFF as functions of the energy fractions $z_1$ and $z_2$, 
integrated over all the other kinematical  
variables of the produced hadron pair, including their 
invariant mass $M_h$. However, the largest amount of experimental information
related to DiFF consists of invariant mass 
spectra of hadron pairs produced in $e^+e^-$ 
annihilation~\cite{Acton:1992sa,Abreu:1992xx,Buskulic:1995gm}, 
Semi-Inclusive Deep-Inelastic Scattering 
(SIDIS)~\cite{Cohen:1982zg,Aubert:1983un,Arneodo:1986tc} 
and proton-proton 
collisions~\cite{Blobel:1973wr,Aguilar-Benitez:1991yy,Adams:2003cc}. In this
paper, using the techniques of 
Jet Calculus~\cite{Konishi:1978yx,Konishi:1979cb}
we deduce the evolution 
equations for DiFF with an explicit dependence on $M_h$. In analogy with 
Ref.~\cite{Camici:1998bg}, we address them as extended Dihadron 
Fragmentation Functions (extDiFF).

DiFF turn out to have important applications in polarization studies, since
they can act as spin-analyzers of the 
fragmenting quark~\cite{Efremov:1992pe}. In particular, the 
transverse polarization $s_\sT$ of 
the fragmenting quark can be related to the azimuthal orientation of the plane
containing the two hadron momenta $P_1$ and $P_2$, through the mixed product
$P_1 \times P_2 \cdot s_\sT$. The strength of this relation is described by
the DiFF $H_1^{\open\, i\to h_1 h_2}$~\cite{Bianconi:1999cd}.
In SIDIS with transversely polarized targets, this DiFF appears in combination
with the transversity distribution
function~\cite{Collins:1994kq,Jaffe:1998hf,Radici:2001na}, thus
providing a way to constrain this elusive partonic distribution 
(for a review on transversity, see Ref.~\cite{Barone:2003fy}). 
The HERMES~\cite{vanderNat:2005yf} and 
COMPASS~\cite{Martin:2007au} collaborations have reported preliminary measurements 
of the induced spin asymmetry (at $\langle Q^2 \rangle\approx2.5$ GeV$^2$). 
In the meanwhile, the BELLE collaboration is planning to perform the extraction of  
the fragmentation function $H_1^{\open}$ in 
$e^+ e^-$ annihilation~\cite{Boer:2003ya}, but at the higher scale 
$\sqrt{s}\approx 10$ GeV~\cite{Hasuko:2003ay}. 
The invariant-mass dependence of this fragmentation function is unknown and
nontrivial, as shown, e.g., by model
calculations~\cite{Jaffe:1998hf,Radici:2001na,Bacchetta:2006un}. 
Therefore, the study of the evolution properties of extDiFF is also timely.
 
The paper is organized as follows. In Sec.~\ref{s:DiFF}, using Jet Calculus we
recover the inhomogeneous evolution equations for DiFF derived in 
Ref.~\cite{deFlorian:2003cg}. In Sec.~\ref{s:extDiFF}, following the same lines
we deduce the evolution equations for the corresponding extDiFF. Finally, in
Sec.~\ref{s:end} we draw our conclusions.

\section{Integrated dihadron fragmentation functions}
\label{s:DiFF}

The cross section at order $O(\alpha_s)$ for the $e^+e^-\to h X$ process, where 
a hadron has momentum $P_h$ and energy fraction $z=2P_h\cdot q/Q^2$ with respect 
to the center-of-mass (cm) energy $Q^2 \equiv q^2$, can be written formally as
\be
\frac{d\sig^h}{dz}=\sum_{i} \sig^i \otimes D_1^{i\to h} \; ,
\label{eq:1hcross}
\ee
where $D_1^{i\to h}$ describes the ``decay" at leading twist of the hard parton 
$i$ into the observed hadron $h$, the sum on $i$ running over all possible 
partons species $i=q,\bar{q},g$. The process-dependent coefficient function
$\sig^i$ can be calculated and regularized in perturbation theory and it is known 
at least at 
$O(\alpha_s)$~\cite{Altarelli:1979kv,Furmanski:1981cw}. 
The $D_1^{i\to h}$ at $O(\alpha_s)$, i.e., which absorbs all the collinear
singularities, can be accurately 
parametrized~\cite{Kretzer:2000yf,Kniehl:2000fe}, except for the $z\to 0$ portion 
of phase space. 

The generalization of Eq.~(\ref{eq:1hcross}) to the process $e^+e^-\to h_1 h_2 X$
is not straightforward, if one wants to cover the whole phase space accessible to 
$(h_1,h_2)$. The differential cross section, again at $O(\alpha_s)$, has been 
recently calculated in Ref.~\cite{deFlorian:2003cg} and reads, with obvious 
notations, 
\be
\frac{d\sig^{h_1,h_2}}{dz_1 dz_2}=\sum_{ij}\sig^{ij}\otimes D_1^{i\to h_1}
\otimes D_1^{j\to h_2}+\sum_{i}\sig^{i}\otimes D_1^{i\to h_1 h_2} \; ,
\label{eq:2hcross}
\ee
where the DiFF $D_1^{i\to h_1 h_2}$ contains information on the fragmentation, 
at leading twist and $O(\alpha_s)$, of the hard parton $i$ directly into the 
observed hadron pair $h_1, h_2$. At order $O(\alpha_s^0)$, the first term of 
Eq.~(\ref{eq:2hcross}) would correspond to the back-to-back emission of a parton 
and an antiparton, eventually fragmenting in the hadrons $h_1$ and $h_2$ 
belonging to two well separated jets. The second term would apply instead to the 
case where the hadron pair is produced very close in phase space and it is detected inside 
the same jet while the other jet is inclusively summed over. However, at order 
$O(\alpha_s)$ a new kind of collinear singularities arises in the partonic cross 
section; it corresponds to the configuration where each hadron is obtained from 
the fragmentation of a single parton, the two partons being almost collinear, 
i.e., with a very small relative transverse momentum $r_\sT$. These $1/r^2_\sT$  
singularities cannot be reabsorbed in each $D_1^{i\to h}$, because they do not 
correspond to the back-to-back configuration. Hence, they must be reabsorbed in 
$D_1^{i\to h_1 h_2}$, thus making the two terms in Eq.~(\ref{eq:2hcross})
indistinguishable~\cite{deFlorian:2003cg}.

As a consequence, after integrating over $r_\sT$ the DiFF must satisfy the 
following evolution equation~\cite{deFlorian:2003cg}
\begin{equation} 
\begin{split} 
\frac{d}{d \ln Q^2} D_1^{i\to h_1 h_2}(z_1,z_2,Q^2) &= 
\frac{\alpha_s(Q^2)}{2\pi}\int_{z_1+z_2}^{1}\frac{du}{u^2} \,
D_1^{j\to h_1 h_2} \left(\frac{z_1}{u},\frac{z_2}{u},Q^2\right) \, P_{ji}(u)
\\
& \quad +\frac{\alpha_s(Q^2)}{2\pi}\int_{z_1}^{1-z_2}
\frac{du}{u(1-u)} \, D_1^{j\to h_1}\left( \frac{z_1}{u},Q^2
\right) D_1^{k\to h_2}\left( \frac{z_2}{1-u},Q^2 \right )\, \hat{P}_{jk}^i(u)\; ,
\label{eq:DD-evo}
\end{split}
\end{equation} 
where here, and in the following, a sum over repeated parton indices is
understood. 
The first term in the right-hand side represents the usual homogeneous evolution
for the DiFF $D_1^{i\to h_1 h_2}$ in complete analogy with the case of 
single-hadron fragmentation: the probability for the parton $i$ to fragment into
the hadrons $h_1, h_2,$ is affected by the probability of emitting a parton $j$
with momentum fraction $u$ through the Altarelli-Parisi splitting vertex
$P_{ji}(u)$~\cite{Altarelli:1977zs}, listed in Eqs.~(\ref{eq:Pqq})--(\ref{eq:Pgg})
of the appendix for convenience. The second term is a new inhomogeneous contribution that 
corresponds to the probability for the parton $i$ to split in the two partons 
$j$ and $k$ with momentum fractions $u$ and $(1-u)$, respectively, each one 
fragmenting in one of the two observed hadrons. 
The $\hat{P}_{jk}^i(u)$ are the Altarelli-Parisi splitting functions
without virtual contributions~\cite{Konishi:1979cb}, again listed in
Eqs.~(\ref{eq:hatPqq})--(\ref{eq:hatPgg}) of the appendix for convenience. 
From this point of view, 
the situation is similar to the DIS case in the target fragmentation region, 
since the DiFF can be conceived as the time-like version of the fracture 
functions in the space-like domain~\cite{Trentadue:1993ka,Graudenz:1994dq}.

In the following, we will make use of Jet Calculus ~\cite{Konishi:1978yx,Konishi:1979cb}
and recover the evolution equation~(\ref{eq:DD-evo}) within this formalism.
We will consider 
the semi-inclusive production of two hadrons,
$h_1$ and $h_2$, belonging to the same jet and neglecting the emission of 
wide-angle hard partons (and related jets). Therefore, we will not perform a 
fixed-order calculation of the $e^+e^-\to h_1 h_2 X$ cross section. Rather, we 
will consider a parton $i$ with a large virtuality $\eps Q^2$ ($0<\eps <1$), 
which fragments in two hadrons $h_1$ and $h_2$ inside the same jet. The 
virtuality can be 
reconstructed from the invariant mass of the jet by a suitable jet-finding 
algorithm~\cite{Catani:1991pm}. The phase-space structure of collinear 
singularities singled out in fixed-order calculations can be translated in Jet 
Calculus as a degeneracy in all possible competing mechanisms, which could 
realize the desidered final state configuration~\cite{Konishi:1978yx,Konishi:1979cb}. 
Thus, the cross section is the sum of all production mechanisms, 
as in Eq.~(\ref{eq:2hcross}). 

\begin{figure}[h]
\includegraphics[width=10cm]{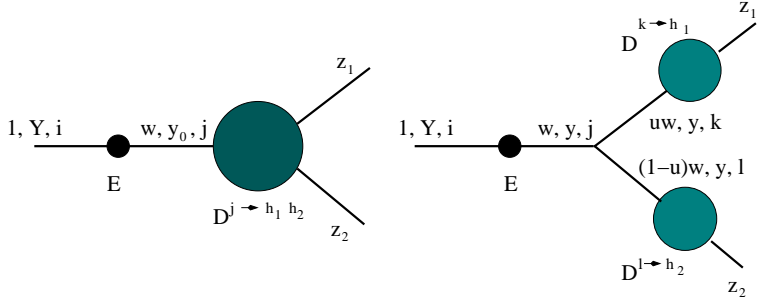}
\caption{Double- and single-hadron fragmentations in 
Eq.~(\protect{\ref{eq:jetcross}}). The momentum fractions are indicated along 
with the scale and the parton indices. The black dot represents the parton 
evolution function $E$.}
\label{fig:jetcross}
\end{figure}

We use $Q^2$ as evolution scale, instead of $\eps Q^2$.
In Leading Logarithmic Approximation (LLA), this substitution induces only 
subleading corrections and thus is fully justified within this approximation. 
Moreover, it is convenient to replace this variable with the evolution variable 
\be
Y=\frac{1}{2\pi \beta_0}\ln \left[ \frac{\alpha_s(\mu_R^2)}{\alpha_s(Q^2)}\right]
\; ,
\label{eq:Y}
\ee
also named the evolution imaginary time~\cite{Konishi:1979cb}. 
In Eq.~(\ref{eq:Y}), $\mu_R^2$ is the renormalization scale and 
$\beta_0=(11 N_c-2 N_f)/(12\pi )$ 
is the one-loop $\beta$ function with $N_c$, $N_f$, the number of colors and
flavors, respectively. In 
LLA, the running of 
$\alpha_s$ is taken into account at one loop by  
\be
\alpha_s(Q^2)=\frac{1}{\beta_0 \ln (Q^2/\Lambda_{QCD}^2)} \; ,
\ee
where $\Lambda_{QCD}^2$ is the infrared scale. Hence, the differential evolution 
length is just
\be
dY = \frac{\alpha_s(Q^2)}{2\pi}\frac{dQ^2}{Q^2} \; .
\label{eq:dY}
\ee
Let us define the variable  $y$ as 
\be
y=\frac{1}{2\pi \beta_0}\ln \left[ \frac{\alpha_s(Q_0^2)}{\alpha_s(Q^2)}\right] 
\; ,
\label{eq:y}
\ee
with $Q_0^2$ and $Q^2$ two arbitrary scales, 
and introduce the perturbative parton-to-parton time-like evolution function 
$E_j^i(x,y)$, which  
expresses the probability of finding a parton 
$j$ at the scale $Q_0^2$ with a momentum fraction $x$ of the 
parent parton $i$ at the scale $Q^2$. The function $E_j^i(x,y)$ can be shown to 
satisfy standard evolution equations~\cite{Konishi:1978yx,Konishi:1979cb}
\be
\frac{d}{dy}E_j^i(x,y)=
\int_x^1 \frac{du}{u} E_j^k 
\left( \frac{x}{u},y \right) P_{ki}(u) \; ,
\label{eq:E-evo}
\ee
that can be iteratively solved by using the initial condition
\be
E_j^i(x,y)|_{y=0}=\delta_{ji}\delta(1-x) \; , 
\label{eq:E0}
\ee
with $\delta_{ji}$ the Kronecker symbol. The $E_j^i(x,y)$ resums 
leading logarithms of the type $\alpha_s^n \ln^n (Q^2/Q_0^2)$, 
which show up in the collinear limit of perturbative calculations at the
partonic level.  
As a crosscheck, we
can expand Eq.~(\ref{eq:E-evo}) at order $O(\alpha_s)$ with the initial
condition~(\ref{eq:E0}); neglecting for simplicity the running of $\alpha_s$ in
Eq.~(\ref{eq:y}), we get 
\be
E_j^i(x,y)\equiv E_j^i(x,Q_0^2,Q^2)\approx\delta_{ji}\delta(1-x)+\frac{\alpha_s}{2\pi}
\, P_{ji}(x) \ln \frac{Q^2}{Q_0^2} \; .
\label{eq:E1}
\ee 
Leading logarithmic contributions
are therefore automatically accounted for at all orders 
through the function $E$. 
Consider now 
the fragmentation process $i\rightarrow h_1 h_2 X$ 
where $h_1$ and $h_2$ are detected within the same jet and 
with relative transverse momentum $R_\sT^2 \lesssim Q^2$.
The corresponding cross-section, normalized to the jet-cross section, can be 
written as
\begin{equation} 
\begin{split} 
\frac{1}{\sig_{jet}}\frac{d \sig^{i \to h_1 h_2}}{dz_1 dz_2} &\equiv 
D_1^{i\to h_1 h_2}(z_1,z_2,Y) =D_{1,A}^{i\to h_1 h_2}(z_1,z_2,Y) +
D_{1,B}^{i\to h_1 h_2}(z_1,z_2,Y) \\
&= \int_{z_1+z_2}^{1}\frac{dw}{w^2}\, D_1^{j\to h_1 h_2} 
\left( \frac{z_1}{w},\frac{z_2}{w},y_0 \right) E_j^i(w,Y-y_0)  \\
&\quad + \int_{y_0}^{Y}dy \int_{z_1+z_2}^{1}\frac{dw}{w^2}\, 
\int_{\frac{z_1}{w}}^{1-\frac{z_2}{w}} \frac{du}{u(1-u)} \hat{P}_{lk}^j(u) E_j^i(w,Y-y) 
D_1^{k\to h_1}\left( \frac{z_1}{wu},y\right) D_1^{l\to h_2}\left( 
\frac{z_2}{w(1-u)},y\right).
\label{eq:jetcross}
\end{split}
\end{equation}
The first term ``$A$" is the convolution of the DiFF at some arbitrary (but still 
perturbative) factorization scale $y_0$ with the parton-to-parton evolution 
function $E_j^i\;$ . The second term ``$B$", 
instead, represents the two separate single-hadron fragmentations, integrated 
over all possible generic intermediate scales $y$ at which the branching 
$j\to kl$ at partonic level might occur. 
Integration limits in  Eq.(\ref{eq:jetcross}) are fixed by momentum conservation. 
Both terms are depicted in Fig.~\ref{fig:jetcross}.
In order to recover Eq.(\ref{eq:DD-evo}), we now take the derivative 
of Eq.~(\ref{eq:jetcross}) with respect to the variable 
$Y$. Using Eq.~(\ref{eq:E-evo}) and the definition of $D_{1,A}^{i\to h_1 h_2}$, 
we get for the ``$A$" term
\be 
\begin{split} 
\frac{d}{d Y} D_{1,A}^{i\to h_1 h_2}(z_1,z_2,Y) &=
\int_{z_1+z_2}^{1}\frac{dw}{w^2} \, 
D_1^{j\to h_1 h_2}\left( \frac{z_1}{w},\frac{z_2}{w},y_0\right ) \, \frac{d}{dY}
[E_j^i(w,Y-y_0)] \\
&=\int_{z_1+z_2}^1 \frac{du}{u^2} \, D_{1,A}^{k\to h_1 h_2}
\left( \frac{z_1}{u},\frac{z_2}{u},Y \right)\,  P_{ki}(u)\; .
\label{eq:DDA-evo}
\end{split} 
\ee
The derivative  $d/dY$ of  the ``$B$" contribution in Eq.~(\ref{eq:jetcross})
produces two terms, since there is an explicit $Y$-dependence in the 
upper integration limit.  
Using Eq.~(\ref{eq:E0}) and the same procedure as before,
we get
\begin{equation} 
\begin{split} 
\frac{d}{dY} & D_{1,B}^{i\to h_1 h_2}(z_1,z_2,Y) =
\int_{z_1+z_2}^{1}\frac{dw}{w^2} \, \delta_{ji} \delta(1-w)
\int_{\frac{z_1}{w}}^{1-\frac{z_2}{w}} \frac{du}{u(1-u)} \, \hat{P}_{lk}^j(u)
D_1^{k\to h_1}\left( \frac{z_1}{wu},Y\right)
D_1^{l\to h_2}\left( \frac{z_2}{w(1-u)},Y\right) \\
& 
\quad \qquad +\int_{y_0}^{Y}dy \int_{z_1+z_2}^{1}\frac{dw}{w^2} 
\int_{\frac{z_1}{w}}^{1-\frac{z_2}{w}} \frac{du}{u(1-u)} \, \hat{P}_{lk}^j(u) \,
\frac{d}{dY} E_j^i(w,Y-y)\, D_1^{k\to h_1}\left( \frac{z_1}{wu},y\right)
D_1^{l\to h_2}\left( \frac{z_2}{w(1-u)},y\right)  \\
& 
\quad =\int_{z_1}^{1-z_2} \frac{du}{u(1-u)} \, 
\hat{P}_{lk}^i(u) D_1^{k\to h_1}\left( \frac{z_1}{u},Y\right)
D_1^{l\to h_2}\left( \frac{z_2}{1-u},Y\right) + 
\int_{z_1+z_2}^{1}\frac{du}{u^2} \, P_{ki}(u) D_{1,B}^{k\to h_1 h_2}
\left( \frac{z_1}{u},\frac{z_2}{u},Y\right).
\label{eq:DDB-evo}
\end{split} 
\end{equation} 
Summing up Eqs.~(\ref{eq:DDA-evo}) and~(\ref{eq:DDB-evo}), we get
\begin{equation} 
\begin{split} 
\frac{d}{dY} D_1^{i\to h_1 h_2}(z_1,z_2,Y) &= 
\int_{z_1+z_2}^1 \frac{du}{u^2} \, D_1^{k\to h_1 h_2}
\left( \frac{z_1}{u},\frac{z_2}{u},Y \right)\, P_{ki}(u) \\
&\quad + \int_{z_1}^{1-z_2} \frac{du}{u(1-u)} \,
D_1^{k\to h_1}\left( \frac{z_1}{u},Y\right)
D_1^{l\to h_2}\left( \frac{z_2}{1-u},Y\right) \, \hat{P}_{lk}^i(u)\; ,
\label{eq:DD-evo2}
\end{split}
\end{equation} 
which is exactly Eq.~(\ref{eq:DD-evo}), after changing the scaling variable $Y$ 
back to the more familiar $Q^2$.
This derivation is useful to adjust the formalism of Jet Calculus from 
Refs.~\cite{Konishi:1978yx,Konishi:1979cb} to the calculation of the 
$e^+e^-\to h_1 h_2 X$ cross section at $O(\alpha_s)$ from 
Ref.~\cite{deFlorian:2003cg}. However, we want to stress again that even if the 
two expressions are formally identical, they have been derived from rather 
different approaches. Eq.~(\ref{eq:DD-evo}) from Ref.~\cite{deFlorian:2003cg}
applies to the full phase space for the production of two hadrons. 
Instead, Eq.~(\ref{eq:DD-evo2}) gives the evolution of the parton $i$ in a jet 
within which we identify the two detected hadrons $h_1$ and $h_2$; thus, it is 
valid only in the portion of phase space defined by the jet. We recall that if
$Q^2$ is the cm energy of the $e^+ e^-$ annihilation, the event is characterized
by a certain number of jets resulting also from a large-angle hard parton
emission; this is not included in the LLA used here.
Therefore, the evolution scale in Eq.~(\ref{eq:DD-evo2}) must be intended as the
invariant mass of the considered jet, i.e., $\eps Q^2$ with $0<\eps <1$.

\section{Extended dihadron fragmentation functions}
\label{s:extDiFF}

In the previous section, we considered the inclusive production of two hadrons 
$h_1$ and $h_2$ inside the same jet, summing up all possible values of their 
invariant mass $M_h$. 
If the process starts from the hard scale $Q^2$ of the
fragmenting parton (or, equivalently, $\eps Q^2$ in the case of the jet),
there is no intermediate scale
that allows to distinguish the two contributions in Eq.~(\ref{eq:jetcross}): the
$y_0$ in the ``$A$" term is arbitrary, and the scale $y$ for the partonic
branching in the ``$B$" term is summed over. 

However, most of the 
experimental information on unpolarized DiFF consists of invariant 
mass spectra of hadron 
pairs~\cite{Acton:1992sa,Abreu:1992xx,Buskulic:1995gm,Cohen:1982zg,Aubert:1983un,Arneodo:1986tc,Blobel:1973wr,Aguilar-Benitez:1991yy,Adams:2003cc}.
In addition, effects related to the
partial-wave expansion of DiFF~\cite{Bacchetta:2002ux}
are best explored when the latter explicitly depend on $M_h^2$.
Hence, in this section we will address the evolution equations for DiFF at the
fixed scale $M_h^2$. We will indicate these objects as extended Dihadron
Fragmentation Functions (extDiFF), as the time-like analogue of the extended 
fracture functions that were introduced in Ref.~\cite{Camici:1998bg} for the
space-like SIDIS in the target fragmentation region.

The dependence of the extDiFF upon $M_h^2$ can be easily mapped into  
$R_\sT^2$, the square of the relative transverse momentum of the hadron pair. 
In fact, it results~\cite{Bacchetta:2002ux}
\be
R_\sT^2 \equiv \frac{(P_{1\sT} - P_{2\sT})^2}{4} =  
\frac{z_1 z_2}{z_1+z_2}\left[\frac{M_h^2}{z_1+z_2}
         - \frac{M_1^2}{z_1} - \frac{M_2^2}{z_2} \right] \; ,
\label{eq:Mh2RT}
\ee
with $M_1$ and $M_2$ the masses of the hadrons $h_1$ and $h_2$, respectively. 
From the first line of Eq.~(\ref{eq:jetcross}), we get the obvious definition
\begin{equation}
\begin{split} 
\frac{1}{\sig_{jet}}\frac{d \sig^{i \to h_1 h_2}}{dz_1 dz_2} &\equiv 
D_1^{i\to h_1 h_2}(z_1,z_2,Y) =D_{1,A}^{i\to h_1 h_2}(z_1,z_2,Y) +
D_{1,B}^{i\to h_1 h_2}(z_1,z_2,Y) \\
&\equiv \int dR_\sT^2 \frac{1}{\sig_{jet}}
\frac{d \sig^{i \to h_1 h_2}}{dz_1 dz_2 dR_\sT^2} =
\int dR_\sT^2\, D_{1,A}^{i\to h_1 h_2}(z_1,z_2,R_\sT^2,Y) + 
\int dR_\sT^2\, D_{1,B}^{i\to h_1 h_2}(z_1,z_2,R_\sT^2,Y) \; .
\label{eq:extDiFF-1}
\end{split}
\end{equation}
The inhomogeneous ``$B$" term describes the time-like branching of parton $i$ in 
two partons $k$ and $l$ with transverse relative momentum $r_\sT$, eventually
fragmenting in the two hadrons $h_1$ and $h_2$ with transverse relative momentum
$R_\sT$. If the $R_\sT^2$ scale is fixed and in the perturbative regime, the scale
at the partonic branching is no longer arbitrary as in
Eq.~(\ref{eq:jetcross})~\cite{Konishi:1978yx,Konishi:1979cb}. In fact, if $u$ and
$(1-u)$ are the fractional momenta of partons $k$ and $l$, the parton
virtualities are related by 
\be
k_i^2=\frac{k_k^2}{u}+\frac{k_l^2}{1-u}+\frac{r_\sT^2}{4u(1-u)} \; .
\label{eq:branch}
\ee
Hence, fixing $r_\sT$ at the partonic level determines in turn the branching scale
$k_i^2$. At the 
hadronic level, this is not guaranteed and some assumptions must be
made. We will suppose that in the fragmentations $k\to h_1$ and $l\to h_2$ the
parton virtualities are negligible, i.e., $k_k^2 \simeq k_l^2 \simeq 0$, meaning 
that, once the branching $i\to k l$ has occurred, both the perturbative and 
nonperturbative transverse momenta generated in the fragmentation of the partons 
$k$ and $l$ are negligible (incidentally, perturbatively generated transverse 
momenta can be taken into account by using time-like evolution equations depending
on transverse momentum~\cite{Ceccopieri:2005zz}). Consequently, the transverse
relative momentum $r_\sT$ between $k$ and $l$ should not be substantially altered 
in the fragmentation, implying 
\be
k_i^2 \approx r_\sT^2 \approx R_\sT^2 \; .
\label{eq:branch2}
\ee
Corrections to the above relation affect our final result only at subleading 
level. Instead, the above assumption is also consistent with the approximation 
we are working with. Leading logarithms are 
known to manifest themselves when the transversa momenta of the emitted partons 
are strongly ordered along the ladder, e.g.,  
$Q^2\gg r_{T,1}^2 \gg \hdots \gg r_{T,n}^2 \gg Q_0^2$. 
Hadron pairs with large relative transverse momentum $R_\sT$
are thus produced earlier in the imaginary time $Y$ 
than hadron pairs with small $R_\sT$, as appropriate for time-like kinematics.

If $R_\sT^2$ is thus in the perturbative domain, 
in analogy with Eq.~(\ref{eq:Y}), we can define the  variable $y_\sT$ as
\be
y_\sT=\frac{1}{2\pi \beta_0}\ln \left[ 
\frac{\alpha_s(\mu_R^2)}{\alpha_s(R_\sT^2)}\right] \; ,
\label{eq:yT}
\ee
or, in differential form,
\be
\frac{d}{dR_\sT^2}=\frac{\alpha_s(R_\sT^2)}{2\pi R_\sT^2} \frac{d}{dy_\sT} \; .
\label{eq:dyT}
\ee
Since the scale at which the branching occurs is fixed by  $R_\sT^2$, from Eqs.~(\ref{eq:jetcross}), (\ref{eq:extDiFF-1}) and (\ref{eq:dyT}), 
we obtain
\begin{equation} 
\begin{split} 
\lefteqn{D_{1,B}^{i\to h_1 h_2}(z_1,z_2,R_\sT^2,Y)} 
\\
&= 
\frac{\alpha_s(R_\sT^2)}{2\pi R_\sT^2} \frac{d}{dy_\sT} 
\int_{y_0}^{Y}dy \int_{z_1+z_2}^{1}\frac{dw}{w^2}\,  
\int_{\frac{z_1}{w}}^{1-\frac{z_2}{w}} \frac{du}{u(1-u)} \hat{P}_{lk}^j(u) \, 
E_j^i(w,Y-y) \, D_1^{k\to h_1}\left( \frac{z_1}{wu},y\right) 
D_1^{l\to h_2}\left( \frac{z_2}{w(1-u)},y\right) 
 \\
&= \frac{\alpha_s(R_\sT^2)}{2\pi R_\sT^2} \int_{z_1+z_2}^{1}\frac{dw}{w^2}\, 
\int_{\frac{z_1}{w}}^{1-\frac{z_2}{w}} \frac{du}{u(1-u)} 
\hat{P}_{lk}^j(u) \,E_j^i(w,Y-y_\sT)\, D_1^{k\to h_1}\left(
  \frac{z_1}{wu},y_\sT\right) 
D_1^{l\to h_2}\left( \frac{z_2}{w(1-u)},y_\sT\right).
\label{eq:extDiFF-B}
\end{split} 
\end{equation} 

If the scale $R_\sT^2$ is fixed in the nonperturbative regime, the above
arguments leading to Eq.~(\ref{eq:dyT}) do not apply. This is the case for the
homogeneous ``$A$" term in Eq.~(\ref{eq:extDiFF-1}), which describes the direct
fragmentation of parton $i$ in the two hadrons $h_1$ and $h_2$: the virtuality $k_i^2$
of the parent parton cannot be reconstructed from $R_\sT^2$ and it
is set to the arbitrary factorization scale $Q_0^2$ (or, in our notations, $y_0$).
From Eqs.~(\ref{eq:jetcross}) and (\ref{eq:extDiFF-1}), we simply get
\be
D_{1,A}^{i\to h_1 h_2}(z_1,z_2,R_\sT^2,Y) = \int_{z_1+z_2}^{1}\frac{dw}{w^2}\, 
D_{1,A}^{j\to h_1 h_2}\left( \frac{z_1}{w},\frac{z_2}{w},R_\sT^2,
y_0\right)\, E_j^i(w,Y-y_0) \;.
\label{eq:extDiFF-A}
\ee
Summing up Eqs.~(\ref{eq:extDiFF-A}) and (\ref{eq:extDiFF-B}), and providing each term 
with an extra step function to separate the two different kinematical
regimes, we get the complete expression for the extDiFF at LLA:
\begin{equation} 
\begin{split} 
&D_1^{i\to h_1 h_2}(z_1,z_2,R_\sT^2,Y) = 
D_{1,A}^{i\to h_1 h_2}(z_1,z_2,R_\sT^2,Y) + 
D_{1,B}^{i\to h_1 h_2}(z_1,z_2,R_\sT^2,Y) \\
& \quad =\int_{z_1+z_2}^{1}\frac{dw}{w^2}\, 
D_{1,A}^{j\to h_1 h_2}\left( \frac{z_1}{w},\frac{z_2}{w},R_\sT^2,y_0\right) 
E_j^i(w,Y-y_0) \, \theta(y_0-y_T) \\
& \quad \quad  + \frac{\alpha_s(R_\sT^2)}{2\pi R_\sT^2} \int_{z_1+z_2}^{1}
\frac{dw}{w^2}\, \int_{\frac{z_1}{w}}^{1-\frac{z_2}{w}} 
\frac{du}{u(1-u)} \hat{P}_{lk}^j(u)  E_j^i(w,Y-y_\sT)
D_1^{k\to h_1}\left( \frac{z_1}{wu},y_\sT\right) 
D_1^{l\to h_2}\left( \frac{z_2}{w(1-u)},y_\sT\right) \, \theta(y_T-y_0) .
\label{eq:extDiFF-2}
\end{split} 
\end{equation} 
Note that, despite the presence of the step functions, 
the separation between the two regimes is still arbitrary,  
since it depends on $y_0$ which is itself arbitrary.
The evolution equations for the extDiFF can be obtained, in analogy with
Eq.~(\ref{eq:DD-evo2}), by taking the derivative with respect to $Y$ [or, 
equivalently, $Q^2$ via Eq.~(\ref{eq:dY})]. By using Eq.~(\ref{eq:E-evo})
and the definition~(\ref{eq:extDiFF-2}) of the extDiFF themselves, we get
\be
\frac{d}{d \ln Q^2}D_1^{i\to h_1 h_2}(z_1,z_2,R_\sT^2,Q^2)=
\frac{\alpha_s(Q^2)}{2\pi}\int_{z_1+z_2}^{1}\frac{du}{u^2}\,
D_1^{j\to h_1 h_2}\left( \frac{z_1}{u},\frac{z_2}{u},R_\sT^2,Q^2\right) P_{ji}(u) \; .
\label{eq:extDiFF-evo}
\ee

We explicitly checked that by integrating Eq.~(\ref{eq:extDiFF-evo}) upon 
$R_\sT^2$ we recover Eq.~(\ref{eq:DD-evo}). 
In conclusion, if the hadron pair is inclusively produced in the same jet at fixed
transverse relative momentum $R_\sT$ (or, equivalently, fixed invariant mass
$M_h$), the evolution equations for the extDiFF are of the standard homogeneous
type. The explicit dependence on this new scale breaks the degeneracy of the two 
production mechanisms described in the previous section. The same arguments 
apply to the SIDIS target fragmentation region where extended fracture functions, 
explicitly depending upon the invariant momentum transfer $t$ between 
the incoming and outgoing hadron, satisfy a homogeneous evolution 
equation~\cite{Camici:1998bg}. 
On the basis of Eq.~(\ref{eq:extDiFF-evo}), we argue that the cross section at
order $O(\alpha_s)$ for the inclusive production of the two hadrons $h_1, h_2,$
inside the same jet and with invariant mass $M_h$, can be expressed in the
factorized form
\be
\frac{d \sig^{i \to h_1 h_2}}{dz_1 dz_2 dR_\sT^2}=
\sum_{i}\, \sig^{i}\otimes D_1^{i\to h_1 h_2}(R_\sT^2,Q^2)  \; ,
\label{eq:factorization}
\ee
where $\sig^{i}$ are the same coefficient functions found in the single-hadron 
inclusive cross section of Eq.~(\ref{eq:1hcross}). In our above derivation, we
used the techniques of Jet Calculus~\cite{Konishi:1978yx,Konishi:1979cb}, where 
the factorization of collinear singularities can be automatically accommodated 
through the use of the parton-to-parton evolution function $E$. 
Exchanges of soft particles are, however, not accounted for.

Eq.~(\ref{eq:extDiFF-evo}) can be conveniently diagonalized using a double Mellin
transformation. We define 
\be
D_{n,m}^{i\to h_1 h_2}(R_\sT^2,Q^2)=\int_0^{1} dz_1 \int_0^{1-z_1} dz_2
\, z_1^{n-1} \, z_2^{m-1} D_1^{i\to h_1 h_2}(z_1,z_2,R_\sT^2,Q^2) \; ,
\label{eq:Mellin}
\ee
and the anomalous dimension
\be
A_j^i(n+m)=\int_0^1 du \, P_{ji}(u) \, u^{m+n-2} \; .
\label{eq:anomdim}
\ee
With simple algebra manipulations, it is easy to verify that 
\be
\frac{d}{d \ln Q^2}D_{n,m}^{i\to h_1 h_2}(R_\sT^2,Q^2)=
\frac{\alpha_s(Q^2)}{2\pi}\, D_{n,m}^{j\to h_1 h_2}(R_\sT^2,Q^2) \, A_j^i(n+m)\; .
\label{eq:extDiFF-evo2}
\ee

The above results can be extended also to polarized extDiFF, in particular to 
the only one surviving when the hadron pair is collinear, i.e.,  
$H_1^{\open\,i\to h_1 h_2}$~\cite{Bianconi:1999cd,Bacchetta:2002ux}. 
The evolution equations for this function have the same form of the
unpolarized case, namely
\be
\frac{d}{d \ln Q^2}H_1^{\open\,i\to h_1 h_2}(z_1,z_2,R_\sT^2,Q^2)=
\frac{\alpha_s(Q^2)}{2\pi}\int_{z_1+z_2}^{1}\frac{du}{u^2}\,
H_1^{\open\,j\to h_1 h_2}\left( \frac{z_1}{u},\frac{z_2}{u},R_\sT^2,Q^2\right)
\delta P_{ji}(u) \; ,
\label{eq:h1ang-evo}
\ee
where the splitting functions $\delta P_{ji}$ for a
transversely polarized fragmenting parton are used~\cite{Stratmann:2001pt}. 
The $\delta P_{ji}$
are listed in Eqs.~(\ref{eq:deltaPqq})--(\ref{eq:deltaPgg}) 
of the appendix. 

These evolution equations can be conveniently used for phenomenological analyses,
since they can connect experimental data taken at different energies.

\section{Conclusions}
\label{s:end}

We have shown that in leading logarithm approximation the so-called extended 
Dihadron Fragmentation Functions (extDiFF), describing the inclusive production 
of two hadrons inside the same jet at fixed invariant mass $M_h$, satisfy 
evolution equations of the same homogeneous type as in the single-hadron 
fragmentation case. We stress that the explicit dependence on the scale $M_h^2$ 
is required to break the degeneracy at $O(\alpha_s)$ between the fragmentation 
from a single parton or after the branching in two collinear partons. 
While the first contribution pertains to 
the nonperturbative regime, in the 
latter the transverse relative momentum $R_\sT$ of the two hadrons can be traced 
back to the transverse relative momentum of the two collinear partons after the 
branching, and, ultimately, to the hard scale of the originating parton. The 
analysis of the corresponding contribution to extDiFF shows that the dependence 
on this perturbative scale can be predicted. 

In our derivation, we used the techniques of Jet 
Calculus~\cite{Konishi:1978yx,Konishi:1979cb}. Factorization of collinear 
singularities can be automatically accommodated through the use of the 
parton-to-parton evolution function $E$. 
On the basis of the simple result for
the evolution equations of extDiFF, we argue that the cross section at order 
$O(\alpha_s)$ for the inclusive production of the two hadrons $h_1, h_2,$
inside the same jet and with invariant mass $M_h$, can be expressed in a 
factorized form involving the same coefficient functions as in the single-hadron
inclusive cross section. A complete proof of this statement 
would require however the inclusion of soft particle exchanges,
which are not accounted for in Jet Calculus approach. 

Evolution equations can be extended also to polarized extDiFF, in 
particular to $H_1^{\open\,i\to h_1 h_2}$~\cite{Bianconi:1999cd}. 
Such fragmentation function can be extracted from 
$e^+ e^-$ annihilation~\cite{Boer:2003ya}, evolved to the scale of semi-inclusive 
deep inelastic scattering measurements and allow the extraction of the transversity 
distribution function~\cite{Collins:1994kq,Jaffe:1998hf,Radici:2001na}.


\begin{acknowledgments}
F.A.C. would like to thank the DESY Theory Group and the Universiy of Parma for 
financial support. This work is part of the European Integrated Infrastructure 
Initiative in Hadron Physics project under the contract number RII3-CT-2004-506078.
\end{acknowledgments}


\appendix*
\section{Splitting functions}
\label{s:app}
The unpolarized leading-order splitting functions $P(u)$ 
read~\cite{Altarelli:1977zs}
\bea
\label{eq:Pqq}
P_{qq}(u)&=&C_F \left[ 2 \left( \frac{1}{1-u} \right)_+ 
+\frac{3}{2}\,\delta (1-u) -1-u \right]\, ,  \\
P_{qg}(u)&=&T_R [1-2u+2u^2]\, , \\
P_{gq}(u)&=&C_F \left[ \frac{2}{u}-2+u \right]\, ,  \\
P_{gg}(u)&=&2C_A  \left[ \left( \frac{1}{1-u} \right)_+ + \frac{1}{u}+u(1-u)-2
\right] + \frac{11 C_A - 2 N_f }{6} \, \delta (1-u) \, , 
\label{eq:Pgg}
\eea
where $N_f$ is the number of flavors,
$C_A = 3$,
$C_F = 4/3$, $T_R = 1/2$; the $``+"$ prescription is defined as usual by
\be
\int_0^1 dz\, f(z) [g(z)]_+ = \int_0^1 dz\, g(z)\, [f(z)-f(1)]\; .
\label{eq:plusdistr}
\ee
The unpolarized leading-order splitting functions $\hat{P}$ are
readily obtained from the previous ones by dropping virtual contributions 
at the endpoint. In our notations, they read:
\bea
\label{eq:hatPqq}
\hat{P}_{gq}^q(u)&=&C_F \left[  \frac{2}{1-u} -1-u \right]\, , \\
\hat{P}_{q\bar{q}}^g(u)&=&T_R [1-2u+2u^2]\, , \\
\hat{P}_{qg}^q(u)&=&C_F \left[ \frac{2}{u}-2+u \right]\, , \\
\hat{P}_{gg}^g(u)&=&2C_A  \left[ \frac{1}{1-u} + \frac{1}{u}+u(1-u)-2
\right]\, . 
\label{eq:hatPgg}
\eea
The transversely polarized 
leading-order splitting functions $\delta P$ read~\cite{Artru:1990zv,Stratmann:2001pt} 
\bea
\label{eq:deltaPqq}
\delta P_{qq}(u)&=&C_F \left[ 2 \left( \frac{1}{1-u} \right)_+ 
+\frac{3}{2}\delta (1-u)-2 \right]\, , \\
\delta P_{qg}(u)&=&0 \, ,  \\
\delta P_{gq}(u)&=&0 \, ,  \\
\delta P_{gg}(u)&=&2C_A  \left[ \left( \frac{1}{1-u} \right)_+ -1\right]
+ \frac{11 C_A - 2 N_f }{6}\,  \delta(1-u)\, .
\label{eq:deltaPgg}
\eea
Due to angular momentum conservation, there is no mixing between quarks and 
gluons. The last splitting function can be used to evolve the DiFF for
linearly polarized gluons, $\delta\hat{G}^{\open}$~\cite{Bacchetta:2004it}.

\bibliographystyle{apsrev}

\end{document}